\documentclass[prd, amsfonts, twocolumn, nofootinbib, showpacs]{revtex4-2}
\usepackage{graphicx, epsfig}
\usepackage{color}
\usepackage{amsmath}
\usepackage{amssymb}
\usepackage{physics}
\newcommand{\be}{\begin{equation}}
\newcommand{\ee}{\end{equation}}
\newcommand{\bea}{\begin{eqnarray}}
\newcommand{\eea}{\end{eqnarray}}

\newcommand{\gapp}{\mathrel{\raise.3ex\hbox{$>$}\mkern-14mu
\lower0.6ex\hbox{$\sim$}}}
\newcommand{\lapp}{\mathrel{\raise.3ex\hbox{$<$}\mkern-14mu
\lower0.6ex\hbox{$\sim$}}}
\def\bbox{{\,\lower0.9pt\vbox{\hrule \hbox{\vrule height 0.2 cm
\hskip 0.2 cm \vrule  height 0.2 cm}\hrule}\,}}

\begin{document}
\title{Searching for small primordial black holes in planets, asteroids and here on Earth}
%\author{}
%\affiliation{ }
\author{De-Chang Dai$^{1,2}$ }
\author{Dejan Stojkovic$^{3}$ }
\affiliation{ $^1$ Department of Physics, national Dong Hwa University, Hualien, Taiwan, Republic of China}
\affiliation{ $^2$ CERCA, Department of Physics, Case Western Reserve University, Cleveland OH 44106-7079}
\affiliation{ $^3$ HEPCOS, Department of Physics, SUNY at Buffalo, Buffalo, NY 14260-1500, USA}
 %%%%%%%%%%%%%%%%%%%%%%%%%%%%%%%%%%%%%%%%%%%%%%%%%%%%%%%

\begin{abstract}
\widetext
Small primordial black holes could be captured by rocky planets or asteroids, consume their liquid cores from inside and leave hollow structures. We calculate the surface density and surface tension of a hollow structure around a black hole and compare them with the density and compressive strength of various materials that appear in nature to find the allowed parameter space. For example, granite or iron can support a hollow asteroid/planetoid/moon of the size of up to $0.1 R_\oplus$. Along the same lines, future civilizations might build spherical structures around black holes to harvest their energy. Using the strongest material that we currently know how to make (multiwall carbon nanotube), to withstand gravity of one solar mass black hole,  the shell must be constructed at distances larger than $10^4 R_\odot$. Alternatively, a fast black hole can leave a narrow tunnel in a solid object while passing through it. For example,  a $10^{22}$g black hole should leave a tunnel with a radius of $0.1$ micron, which is large enough to be seen by an optical microscope. We could look for such micro-tunnels here on Earth in very old rocks, or even glass or other solid structures in very old buildings. While our estimate gives a very small probability of finding such tunnels, looking for them does not require expensive equipment and long preparation, and the payoff might be significant. 
\end{abstract}

%%%%%%%%%%%%%%%%%%%%%%%%%%%%%%%%%%%%%%%%%%%%%%%%%%

\maketitle

{\it Introduction.}
Small primordial black holes (PBHs) are perhaps the most interesting and intriguing relics from the early universe. They can serve as dark matter candidates \cite{Carr:2021bzv}, source primordial gravitational waves \cite{Dong:2015yjs,Ireland:2023zrd}, help in resolving cosmological problems like the domain wall and magnetic monopole problems \cite{Stojkovic:2004hz,Stojkovic:2005zh} etc \cite{Belotsky:2018wph,Belotsky:2014kca,Khlopov:2008qy}.   However, no convincing PBH candidate has ever been observed so far. The purpose of this paper is point out a new avenue for potential discovery of small PBHs - the existence of hollow objects like planets and asteroids that we could search for even with the existing techniques.  Alternatively, we could use large plates or slabs of metal as effective indicators of a potential PBH passage through them. By small PBHs we assume here mainly the black holes in the mass range $10^{17}$g - $10^{24}$g which are still poorly constrained by observations \cite{Carr:2020gox}.

\begin{figure}[h]
   %\centering
\includegraphics[width=3cm]{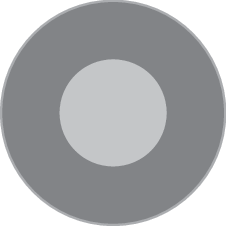}
\caption{ A possible structure of a rocky planet or an asteroid. There is a liquid core (gray) at the center, surrounded by a solid shell (dark) outside. In general, the central core has a higher density than the outside solid shell, so a small black hole can accrete it more efficiently.  
}
\label{core}
\end{figure}

\begin{figure}[h]
   %\centering
\includegraphics[width=8cm]{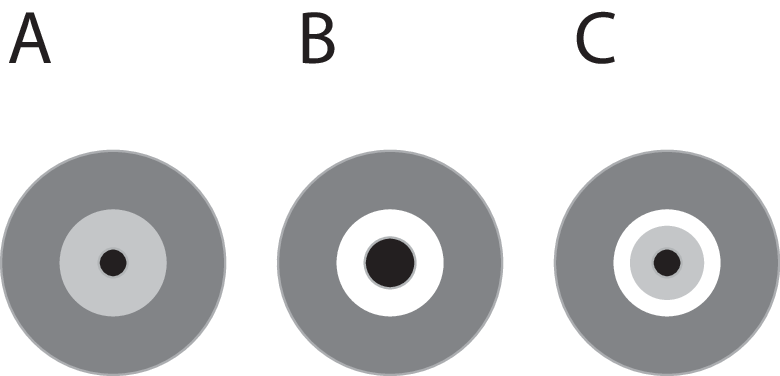}
\caption{Formation of a hollow planet or an asteroid: (A) A planet is formed around a small primordial black hole (or alternatively a planet captures a black hole in its center)  (B) The central core gets slowly absorbed by the black hole. If the outer shell has a strong enough compressive strength, then the shell can support itself leading to a hollow object. (C) If the liquid core becomes solid before it is completely eaten by the black hole, there will exists an empty shell between the outer layer and  central core.   
}
\label{capture}
\end{figure}

It has been recently argued that the main sequence, neutron and dwarf stars can contain small PBHs in their interiors \cite{{Caplan:2023ddo}}. A PBH can be either captured by a star (less likely) or be trapped in the interior during the star formation (more likely). The gas inside these stars could be slowly eaten by such a PBH located inside. We extend this idea to planets and asteroids, which can also be expected to host  PBHs. The capture of a PBH can happen during or after the creation of these objects. If a planet or an asteroid has a liquid central core and an outer solid layer (as shown in Fig.~\ref{core}), then a captured PBH can absorb the liquid core whose density is higher than the density of the outer layer. If the solid layer around the core  is strong enough to support itself, then an hollow planet/asteroid is formed. This process is illustrated in Fig.\ref{capture}.  Any eventual future collisions between this massive shell and other asteroids may separate the shell from the black hole. Thus, such a hollow structure may or may not have a PBH currently in the center. 

The crucial question in this scenario is whether the massive shell can support itself or it will crush under its own tension. To answer this question we will solve a simple general relativistic problem to find the ratio between the surface tension and surface density of the massive shell and compare it with the compressive strengths of the known materials that appear in nature, as well as with the strongest man made materials.  

\begin{figure}[h]
   %\centering
\includegraphics[width=8cm]{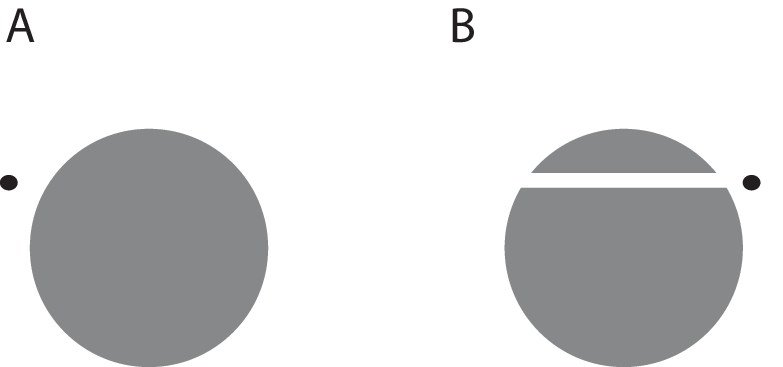}
\caption{ Collision between a fast black hole (black circle) and a solid asteroid (gray circle). (A) before the  black hole - asteroid interaction. (B) as the black hole passes through the asteroid, it leaves a straight tunnel with a cross section comparable to the PBH's Schwarzschild radius. If the composition of the asteroid is strong enough, this tunnel will not promptly disappear.    
 }
\label{cross}
\end{figure}

It is also possible that an object like an asteroid is completely solid without a liquid core. In that case, the interaction with a PBH will not result in an hollow sphere. Since the cross-section of a small PHBs is very small, a fast enough PBH will most likely create a straight tunnel after passing through the asteroid (Fig.~\ref{cross}). Thus, the existence of straight tunnels in an asteroid could also be a sign of an interaction with a PBH. The same effect could allow detection of a PBH here on Earth if we look for sudden appearance of narrow tunnels in metal slabs.

{\it Surface density and tension of a shell around a black hole.~}   
To obtain analytic description of a hollow planet we now solve the following general relativistic problem.   We assume a spherical symmetric spacetime with the metric 
\begin{equation}
ds^2=A(r)dt^2 -B(r)dr^2-r^2 d\theta^2 -r^2 \sin^2 \theta d\phi^2 .
\end{equation}
 At this stage, we suppress the Newton's constant, $G=6.674\times 10^{-11}$ m$^3$kg$^{-1}$s$^{-2}$, and the speed of light, $c=3\times 10^8$m/s, however we will restore them back when needed to go to the SI units.
 
We want to describe a central black hole with mass $M$ surrounded by a massive shell with mass $\delta M$ located at $r=r_0$.  For simplicity, we use a thin shell approximation. The metric function $A(r)$ takes the form  
\begin{equation}
A(r) = 
\begin{cases}
1-\frac{2(M+\delta M)}{r} & \text{for}\ r>r_0\\
\Big(1-\frac{2M}{r}\Big) \frac{1-\frac{2(M+\delta M)}{r_0}}{1-\frac{2M}{r_0}} &\text{for}\ r_0<r 
\end{cases}
\end{equation}
while the metric function $B(r)$ takes the form 
\begin{equation}
B(r) = 
\begin{cases}
\frac{1}{1-\frac{2(M+\delta M)}{r}} & \text{for}\ r>r_0\\
\frac{1}{1-\frac{2M}{r}}  &\text{for}\ r_0<r
\end{cases}
\end{equation}

The corresponding Einstein tensor is 
 
\begin{eqnarray}
G^t_t&=&-\frac{-B+B^2+rB'}{r^2 B^2}\\
G^r_r&=& -\frac{-1+B-rA'}{r^2 A B}\\
G^\theta_\theta&=&-\frac{rBA' {}^2 +2A^2B' +A(rA' B' -2B(A'+rA''))}{4rA^2B^2}\\
G^\theta_\theta&=&G^\phi_\phi .
\end{eqnarray} 

From here, we obtain the surface density, $\lambda$, of the thin shell
\begin{equation}
-8 \pi \lambda=\int_{r_0^- }^{r_0^+} G^t_t \sqrt{B} dr =\frac{2}{ B^{\frac{1}{2}}r} \bigg\rvert_{r_0^- }^{r_0^+}\approx \frac{-2\delta M}{ r_0^2\sqrt{1-2\frac{M}{r_0}}} .
\end{equation}

The surface tension, $\tau$, of the thin shell is
\begin{equation}
-8\pi \tau=\int_{r_0^- }^{r_0^+} G^\theta_\theta \sqrt{B} dr =\frac{1}{rB^{\frac{1}{2}}}+\frac{A'}{2A B^{\frac{1}{2}}} \bigg\rvert_{r_0^- }^{r_0^+}\approx \frac{\frac{M}{r_0}}{(1-\frac{2M}{r_0})^{\frac{3}{2}}}\frac{\delta M}{r_0^2} .
\end{equation}
The ratio between these two parameters is
\begin{equation}
\label{tension-ratio}
\left|\frac{\tau}{\lambda}\right|=\frac{M}{2(r_0 -2M)} .
\end{equation}
This ratio plays the most important role in our discussion. 

In the limit where $r_0\gg 2M$, we recover Newtonian gravity. If $\tau=\lambda/2$, we recover the photon circle (region of space where gravity is so strong that photons are forced to travel in orbits).   

Whether the shell can be supported without crushing depends on the ability of the material to withstand the surface tension. In the other words, the surface tension must be smaller than the compressive strength of the material. The density and compressive strength of common materials found on asteroids and planets are listed in the table \ref{table:nonlin}. For comparison we added the strongest possible man-made material --- multi wall carbon nanotube.

\begin{table}[ht]
\centering % used for centering table
\begin{tabular}{|c| c| c| c|} % centered columns (4 columns)
\hline\hline %inserts double horizontal lines
material & density   & compressive  & maximal  \\ [0.5ex] % inserts table
         & (Mg/m$^3$)& strength (GPa) &  $|\frac{\tau}{\lambda}|$ \\ [0.5ex] % inserts table
%heading
\hline % inserts single horizontal line
Multiwall & 1.8 & 10 to  & $6.2\times 10^{-11}$ to \\ % inserting body of the table
Carbon  nanotube&  & 60 & $3.7\times 10^{-10}$ \\ % inserting body of the table
\hline
Diamond& 3.4& $>$10 & $>3.3 \times 10^{-11}$ \\
\hline
 Quartz & 2.2& $>$1.1 & $>5.6 \times 10^{-12}$ \\
\hline
Iron, Steel & 7.8 to& 0.1 to & $1.4\times 10^{-13}$ to \\
  & 8 &  1 & $1.4\times 10^{-12}$ \\
\hline
Granite  & 2.54 to & $9.65\times 10^{-2}$ to & $4.1\times 10^{-13}$ to \\
  &  2.66 &  $3.1\times 10^{-1}$ & $1.3\times 10^{-12}$ \\
\hline %inserts single line
Concrete  & 2.7  & $17\times 10^{-3}$ to & $7.2\times 10^{-14}$ to \\
  &   &  $70\times 10^{-3}$ & $2.9\times 10^{-13}$ \\
\hline %inserts single line
Ice  & 0.91  & $5\times 10^{-3}$ to & $6.1\times 10^{-14}$ to \\
 (-10 to -20 C) &   &  $25\times 10^{-3}$ & $3.1\times 10^{-13}$ \\
\hline %inserts single line
\end{tabular}
\caption{The density and compressive strength of various materials. The multiwall carbon tube (a man-made material) has the highest compressive strength. Usually, rocky planets and asteroids have a solid outer layer. Quartz, iron, granite and ice are the standard materials of the outer layer.  The data is from the CRC handbook \cite{CRC},  MatWeb (https://www.matweb.com/index.aspx),  \cite{ICE}, and WIKI. Note that compressive strength divided by the product $\rho c^2$, where $\rho$ is the density of the material, is actually $\tau/\lambda$.}  
\label{table:nonlin}
\end{table}

In \cite{Brady:1991np}, it was shown that the stability against radial perturbations becomes an issue only when the shell is very close to the horizon, of the order of the radius of the circular photon orbit. In our case where the black hole is microscopic and the shell macroscopic, stability is not an issue as long as the strength of the material is sufficient to support gravity.  

{\it Planet mass and radius relation.~}
Consider now a rocky planet  (or an asteroid) of mass $M_p$ and radius $R_p$.
The density of the planet, $\rho_p$, is related to its radius as \cite{2014ApJ...783L...6W,2014PNAS..11112655M}
\begin{equation}
\rho_p=2.43+3.39\Big(\frac{R_p}{R_\oplus}\Big) \text{ g/cm$^3$} .
\end{equation}
Here, $M_\oplus=5.9\times 10^{24}$kg and $R_\oplus=6.371\times 10^3$km are the Earth's mass and radius respectively. The planet's mass is related to its radius as 
\begin{equation}
\label{mass-radius}
\frac{M_p}{M_\oplus} = 
\begin{cases}
\Big(\frac{\rho_p}{\rho_\oplus} \Big)\Big( \frac{R_p}{R_\oplus}\Big)^3 \ & \text{for}\ r<1.5R_\oplus\\
2.69\Big(\frac{R_p}{R_\oplus}\Big)^{0.93} &\text{for}\ 1.5R_\oplus<r<4R_\oplus .
\end{cases}
\end{equation}

\begin{figure}[h]
   %\centering
\includegraphics[width=8cm]{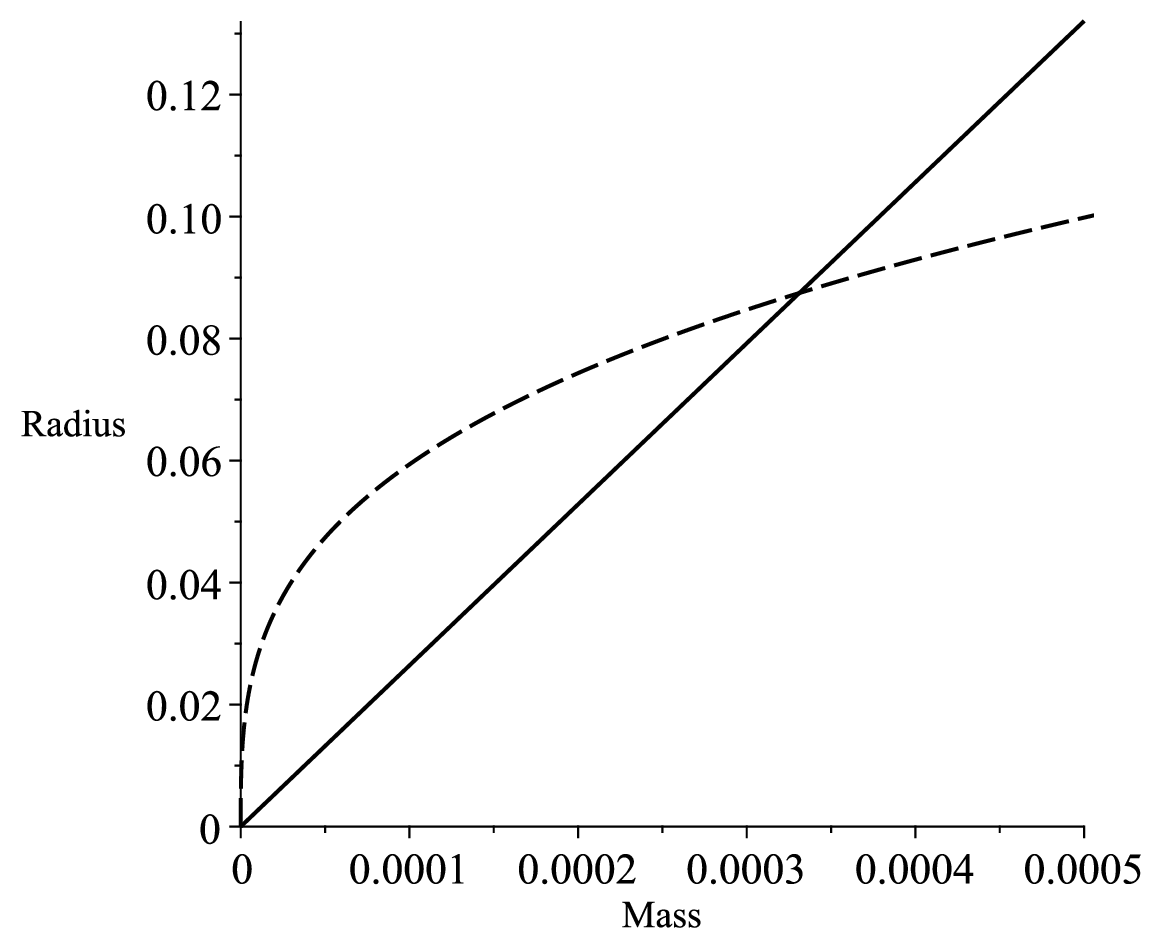}
\caption{We plot the radius of a stable hollow object in units of $R_\oplus$ on y-axis, and its mass in units of $M_\oplus$ on x-axis. We take granite as the main material that the outer layer of an object is made of with $|\tau/\lambda| = 1.3\times 10^{-12}$.  The dashed line is the planet mass-radius relation, Eq.~(\ref{mass-radius}). The solid line is from Eq.~(\ref{tension-ratio}), so the region above the line allows for the hollow shell structure to exist.  For this value of  $|\frac{\tau}{\lambda}|$ it is possible to find a stable hollow planetoid/asteroid/moon of the size up to $0.1 R_\oplus$.
}
\label{radius}
\end{figure}

{\it Parameter space for the hollow structures.~}
We plot the main result in Fig. \ref{radius}.  We pick the value for the ratio  $|\tau/\lambda|=1.3\times 10^{-12}$, which is for example supported by granite or iron.  We plot the planet radius vs. mass relation from Eq.~(\ref{mass-radius}) (dashed curve) and also radius vs. mass relation for the fixed value $|\frac{\tau}{\lambda}|=1.3\times 10^{-12}$ from  Eq.~(\ref{tension-ratio}) (solid line). While $M$ in Eq.~(\ref{tension-ratio}) is the black hole mass, we assume that its mass is smaller than the planet's mass, so the maximum gravity at the outer layer is about the mass of the planet.  Region above the solid line is the allowed region supported by granite or iron. Therefore, the parameter space for a stable hollow object is the portion of the dashed curve located above the solid line. We see that it is possible to find a stable hollow planet (or rather planetoid/asteroid/moon) of the size up to $0.1 R_\oplus$. Several asteroids/planetoids, which have or have ever had a liquid core, were found in this region. First example is Lutetia which has a diameter of about $100$km,  mass of $1.700\times 10^{18}$ kg, and density of  $3.4$  g/cm$^3$. The other one is Vesta which has a diameter of $525.4$ km, mass of $2.590\times 10^{20}$ kg, and density of 3.456 g/cm$^3$. These and similar objects should be the focus of future observations aimed to look for the candidates  in our vicinity whose liquid core was eaten by a primordial black hole. It is interesting that some asteroids like Bennu and Ryugu are possibly hollow \cite{2020SciA....6.3350S,2019Sci...364..268W} because of their low mass density ($\approx 1.2$g/cm$^3$). Though the most likely explanation for their low density is their rubble-pile structure, their discovery and measured properties imply that the search techniques already exist for this type of asteroids.

\begin{figure}[h]
   %\centering
\includegraphics[width=8cm]{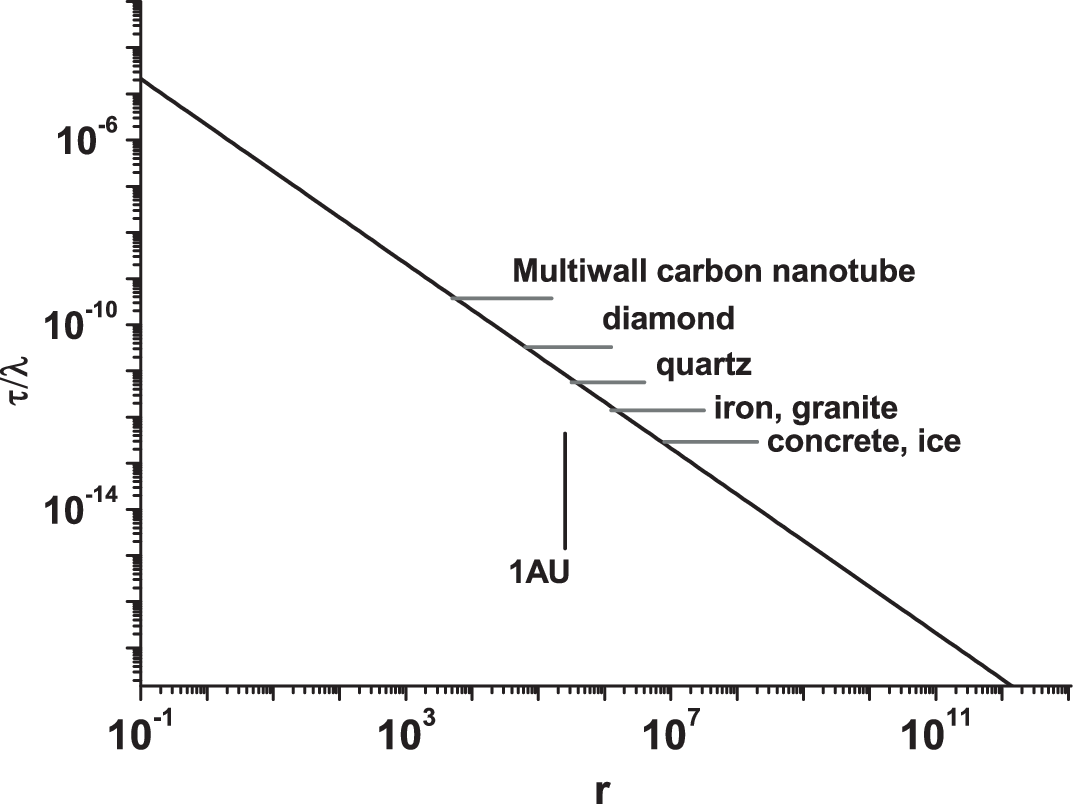}
\caption{ The ratio $|\tau/\lambda|$ of a massive shell as a function of its radius (from  Eq.~(\ref{tension-ratio})). We set the central black hole mass to one solar mass, $1M_\odot$. The horizontal lines represent the maximal value of the ratio $|\tau/\lambda|$ for the corresponding material. The vertical line represent radius of $1AU=1.4\times 10^{11}$km. The x-axis units are $R_\odot=6.96\times 10^5$km. 
 }
\label{sun-radius}
\end{figure}

{\it Structures built around black holes by future civilizations.~}
The calculations we presented here can be used in another technically similar context. 
Namely, it has been argued in the literature that future civilizations may construct a shell outside a rotating or non-rotating black hole in order to harvest energy from it. In the case of a non-rotating black hole one would dump waste and other raw material and harvest pure energy in the form of Hawking radiation. This implies that such a black hole has to be very small in order to achieve a reasonable Hawking energy flux. In the case of a rotating black hole, one would harvest energy emitted from the accretion disk. We can then readily calculate the safe distance at which the harvesting structure must be placed.    
Fig. \ref{sun-radius} shows the required  $|\tau/\lambda|$ to construct a massive shell around a solar mass, $M_\odot=3.955\times 10^{30}$kg, black hole (with a gravitational radius of about $3$km). The strongest material that we currently have is the multiwall carbon nanotube. Even this material cannot withstand this black hole's the gravity at $1 R_\odot$, and the shell must be constructed at distances larger than $10^4 R_\odot$. While our calculations assumed the metric of a non-rotating black hole, we do not expect that the order of magnitude estimate will change significantly for a rotating black hole, especially since this distance is far greater than the size of the ergosphere of the black hole in question. Basically, the structure is in the Newtonian regime, so rotation of the black hole does not change much. However, if the entire structure rotates around a black hole, it can be established closer to the black hole.

{\it Timescales.~}
To verify the validity of the hollow planet scenario proposed here, we now estimate the relevant timescales. The time dependent mass evolution of an accreting black hole (assuming Bondi accretion) is given by \cite{Caplan:2023ddo}
\begin{equation}
M_b (t) =\frac{1}{M_0^{-1}-\frac{\pi \rho G^2 }{c_s^3}t} ,
\end{equation}
where $G$ is the Newton's gravitational constant, $c_s$ is the speed of sound in the accreted material, $\rho$ is density of the material,  $M_b (t) $ is the black hole mass at some moment $t$, while $M_0$ is the initial mass of the seed black hole.   
We can invert this equation to obtain a time interval needed for a complete fluid object (e.g. a star) to be eaten  
\begin{equation}
\tau_B=1.3\text{Myr} \frac{150g/cm^{-3}}{\rho} \frac{M_0}{10^{-12}M_\odot} \Big(\frac{c_s}{550km/s}\Big)^3 ,
\end{equation}
where $M_\odot$ is the solar mass.
We see that this time is at least a Myr (or longer if convection is considered) if the mass of an object is larger than the mass of the seed primordial black hole. 
 However, if a planet contains a central liquid core, the time to consume only the core  could be significantly reduced. For example, a liquid core with a radius of one kilometer has a mass about $3\times 10^{-18}M_\odot$. If the mass of the primordial black hole is $10^{-12}$ to $10^{-10}M_\odot$ (which is within the range of PBHs which are not well constrained by observations),  the time interval would be reduced by a factor of $10^{-6}$ to $10^{-8}$, bringing it down to years or months instead of millions of years. 

{\it Straight tunnels through asteroids or solid material on Earth.~}
In the case of the interaction between a very fast small black hole and a solid asteroid we argued that a narrow tunnel going partially or completely through the asteroid could be formed. 
Since the tunnel is very narrow (compared to the asteroid size),  the material strength requirements to support this tunnel will be less stringent than in the hollow planet/asteroid case. Thus, the results we derived above for the hollow case could be viewed as a very conservative limit for the tunnel case.

It was argued in \cite{Luo:2012pp} that a possible signature of a small PBH transit through Earth are low magnitude earthquakes, while the non-seismic collateral
damage due to the impact would be negligible. Along the lines we presented here, we point out that in fact, a large flat slab of metal (or any other solid material) can serve as a very effective small black hole detector.  We can prepare a polished slab of metal and scan its surface over time. If a straight tunnel suddenly appears,  it could be an evidence of a passage of a small black hole. For example, a $10^{-11} M_\odot$ black hole should leave a tunnel with a radius comparable to its Schwarzschild radius, i.e.  $0.1$ micron. Such a tunnel should be easily detectable with an optical microscope.  It should also be easy to distinguish a small black hole signature from a micrometeorite. A micrometeorite would leave a dent on the slab rather than producing a tunnel.  Also, such a black hole should not be fatal when passing though the human body. While its kinetic energy can be huge, the energy that it can release during the collision with the human body is small. Since the tension of the human tissue is small, it will not tear the tissue apart.

The dark matter density around the sun is $\rho_d=0.43$ GeV/cm$^3$ \cite{Salucci:2010qr}, while the sun's velocity relative to the  Milky Way is $220$km/s. The flux of dark matter partcles (number of partcles per unit area per unit time) is given as 
\begin{equation}
n_d v_d, 
\end{equation}
where $n_d$ is the number density of the particles, while $v_d$ is their velocity.  If all the dark matter is made of PBHs with mass of $m_d=10^{14}$g, then the PBH flux would yield around $\rho_d/m_d v_d \sim 10^{-16}$ crossing events per year through the plate with an area of 1m$^2$. This number is too low for any reasonable real-time observation. However, we could look for straight micro-tunnels in other solid materials like glass or rocks that old buildings are made of, which existed for hundreds and perhaps thousands of years. Ultimately, to maximize the probability of finding them, we could look for these tunnels in rocks which are billions of years old. For example, and a billion year old rock with a cross-section area of $10$m$^2$ should have accumulated $10^{-6}$ crossings so far. While this number is still very small, looking for such tunnels does not require expensive equipment and long preparation. The situation is perhaps similar to the searches for magnetic monopoles. While the theoretical probability of finding a single monopole within our cosmological horizon volume is minuscule (due to inflation), we still search for them since the payoff might be very significant.

{\it Conclusions.}
We proposed here that an interesting new signature of PBHs could be the existence of hollow objects like planetoids, asteroids and moons. Usual materials that these objects are made of can support hollow structures with the radii up to  $0.1 R_\oplus$.  Alternatively, a very fast PBH can leave a narrow tunnel in a solid object while passing through it. In principle, we could look for such micro-tunnels in old rocks, or even glass structures standing for hundreds of years in old buildings. While our estimate gives a very small probability of finding such tunnels,  a very low cost and possible huge payoff should be enough to motivate such a search. Finally, future civilizations might build spherical structures around black holes to harvest their energy. Using the strongest material that we currently know how to make (multiwall carbon nanotube), to withstand gravity of one solar mass black hole,  the shell must be constructed at distances larger than $10^4 R_\odot$.
\begin{acknowledgments}
D.C. Dai is supported by the National Science and Technology Council (under grant no. 111-2112-M-259-016-MY3).    DS is partially supported by the US National Science Foundation, under Grants No.
PHY-2014021 and PHY-2310363.
\end{acknowledgments}

\end{document}